\documentclass[twocolumn, 12pt]{article}
\usepackage[T1]{fontenc}
\usepackage[british]{babel}
\usepackage[utf8x]{inputenc}
\usepackage{hyperref}
\usepackage{graphicx}
\usepackage{amsthm}
\usepackage{amsmath}
\usepackage{verbatim}
\usepackage{amssymb}
\usepackage{attrib}
\usepackage{authblk}

\newcommand{\bra}[1]{\left< #1 \right|}
\newcommand{\ket}[1]{\left| #1 \right>}
\newcommand{\braket}[2]{\left< #1 \middle| #2 \right>}
\newcommand{\ketbra}[2]{\left| #1 \middle>\!\middle< #2 \right|}

\title{Quantum Consensus: an overview}
\author[1]{Marco Marcozzi}
\author[1]{Leonardo Mostarda}
\affil[1]{Computer Science Division, University of Camerino, I-62032 Camerino (MC), Italy}
\date{}

\begin{document}

\twocolumn[
  \begin{@twocolumnfalse}
    \maketitle
    \begin{abstract}
      We review the literature about reaching agreement in quantum networks, also called quantum consensus.
      After a brief introduction to the key feature of quantum computing, allowing the reader with no quantum theory background to have minimal tools to understand, we report a formal definition of quantum consensus and the protocols proposed.
      Proposals are classified according to the quantum feature used to achieve agreement. 
    \end{abstract}
    \vspace{21pt}
  \end{@twocolumnfalse}
  ]

\section*{Introduction}
The advent of Blockchain technology and, more in general, of the Distributed Ledger Technologies (DLTs), has attracted and focused the attention on the study of Distributed systems.
It has been revised the \emph{Consensus problem}, with new proposals being analysed and tested, because of the necessity of agreement on the publication of block, to ensure integrity, and to handle faulty nodes in blockchains~\cite{yaga2019}.\\
Analogously, the increasing interest in quantum computation is leading to a deeper understanding of quantum distributed systems, and eventually it will be implemented a world wide quantum network, already know as Quantum internet~\cite{kimble2008} - the quantum counterpart of the World Wide Web.
In this scenario, the study of quantum distributed systems is fundamental for the correct functioning of such a network~\cite{denchev2008}.
To ensure agreement in a quantum network, thus, it is vital the development of quantum consensus protocols: from a formal definition, to actual proposals.\\

Our paper is structured as follows: an introduction to computational complexity theory will show the importance of quantum computation, then a brief presentation about quantum computing theory will give an insight about quantum formalism to a reader unfamiliar with quantum mechanics, finally an overview about quantum consensus is shown.\\

Computation theory was born and it is developing following the evolution of Information Technology: from the first computers, through the transistors, arriving to the today's supercomputers and the enormous amount of data produced every day~\cite{metropolis2014}.\\
Even though the computational power of our devices nowadays is unbelievable high, if compared with the devices from a few years ago, classical computers seem not to be fit for the next generation of problems in almost any field: chemistry, biology, medical technology, cryptography, optimisation, finance, etc.~\cite{das_top_2020}\\
Because of those and other applications, in last forty years, the possibility to build a computer based on quantum mechanics is focusing the attention and efforts of researchers, governments and companies.
The history of Quantum Computation started in 1980, when Paul Benioff proposed the first quantum mechanical model of a computer~\cite{Benioff1980}.
The next year, Richard Feynman gave a talk at the \emph{First Conference on the Physics of Computation}, in which he stated that quantum mechanical phenomena cannot be efficiently simulated on a classical computer, proposing a basic model for a quantum computer~\cite{Feynman1982}.\\

Since then, a lot of discoveries have been done.
Among the most important ones, there are the description of the first Universal Quantum Turing machine, i.e. the definition of the first universal quantum computer, that can simulate any other quantum Turing machine with at most a polynomial slowdown~\cite{Deutsch1985}, and the groundbreaking discover of the Shor's algorithm, that allows to solve in polynomial time the factoring and the discrete logarithms problem~\cite{Shor1994}.
Shor's algorithm sparked a tremendous interest in quantum computers. The reason lies with the fact that Shor's algorithm can theoretically break many of the cryptosystems in use today~\cite{bernstein2009}.\\

Therefore, the discover of quantum algorithms solving some problems believed to be computationally \emph{hard}, triggered a rephrase of the computational complexity theory.\\
A problem is said to be computationally \emph{easy}, hence it belongs to the computational class \textbf{P}, if it can be solved by a classical computer in polynomial time, respect to the number of bits needed to describe the problem.
A problem is said to be computationally \emph{hard} if the required resources (time) for solving it rises super polynomially fast (often exponentially) with the input size.
Such problems belongs to the computational complexity class \textbf{NP} (non-deterministic polynomial time).
Regarding those problems classes as sets, it is "strongly believed" that \textbf{P}$\subset$\textbf{NP}.
"Strongly believed" means that nowadays it is only know that \textbf{P}$\subseteq$\textbf{NP}, but it is strong the trust in the fact that \textbf{NP} problems cannot be solved in polynomial time by a classical computer.\\
\par
\begin{figure}[ht]
\centering
\includegraphics[width=.45\textwidth]{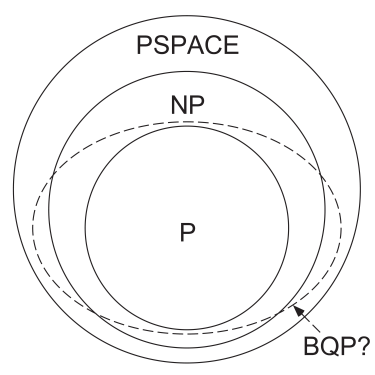}
\caption{The relationship between classical and quantum complexity classes. While it is strongly assumed that \textbf{BQP} is larger than \textbf{P} and encompasses some problems in \textbf{NP}, it remains unclear how the classes are exactly related. Figure from~\cite{nielsen_chuang}.}
\label{fig:complexity}
\end{figure}

In this framework quantum computers play their role.
Shor's algorithm (and other algorithms, e.g. Deutsch's algorithm~\cite{deutsch1992}) demonstrated that certain problems in the \textbf{NP} set could be solved efficiently (i.e. in polynomial time) using a quantum computer.
It leads to a new complexity class, \textbf{BQP} (bounded-error quantum polynomial time), that includes all the problems in \textbf{P} and some problems in \textbf{NP} (it is still not clear how many).\\
In Figure~\ref{fig:complexity} the relationship between the different complexity classes is shown.\\
\par
In this context, scientists started talking about the possible "quantum supremacy" for what concerns computation.
It was conjectured that indeed quantum computers would solve hard problems efficiently, but an experimental proof was needed.
In October 2019, Google and NASA claimed to have achieved and proved the quantum supremacy~\cite{Arute2019},~\cite{Pednault2019}.
The result was impressive, even though there was scepticism, because some researchers pointed out that the problem solved by this quantum annealer (not an universal quantum Turing machine) could have been solved by a classical supercomputer in a comparable amount of time; moreover, since the machine is not a quantum computer, strictly speaking, they believe it is improper to address the achievement of quantum supremacy in this particular context~\cite{Pednault2019}.

\section*{Quantum Computing}
\begin{quote}
    Computers are physical objects, and computations are physical processes.
    \attrib{David Deutsch}
\end{quote}
This sentence reveals a deep connection between information theory, computer science and physics.
Even though information theory can be stated in a completely mathematical fashion, i.e. the physical support used can be omitted, computation theory and algorithms need to rely on a physical theory. It means that, as classical computers uses Maxwell's electromagnetism in a Newtonian framework, a quantum computer needs quantum physics and related phenomena to work.
A brief review, based on~\cite{nielsen_chuang}, about quantum mechanics is presented in the following.\\
\par
Quantum mechanics is an axiomatic theory used as a mathematical framework for the development of physical theories.
It doesn't hold any information about physical laws, but it provides a conceptual connection between the mathematical formalism and the physical world.
As an axiomatic theory, then, quantum mechanics relies on some axioms, or postulates.
For the purposes of this review, those postulates will be not presented rigorously.
Therefore, na\"{i}vely, postulates of quantum mechanics define: the system, as a complex vector space (i.e. an Hilbert space called \emph{state of the system}), how a quantum mechanical state \emph{evolves}, and the \emph{measurement operations}.\\
\par
For computational purposes, the minimum quantum state is represented by a vector in a bi-dimensional complex space.
Such a vector is called \emph{qubit}.
It is important to point out that there exist computational paradigms where a $d$-dimensional Hilbert space is used: in such spaces the unit of information is called \emph{qudit}.\\
A qubit is built on the classical concept of bit and it is the smallest information resource for quantum computers.
While a classical bit can be either 0 or 1, the qubit can be in a \emph{superposition} of both at the same time.
This feature is intrinsic in the properties of the Hilbert space: any combination of vectors in the Hilbert space is a vector that belongs to the same Hilbert space.\\
A convenient way of expressing vectors in an Hilbert space is the Dirac notation, also known as bra-ket notation.
In such notation a vector $v$, called \textit{ket}, is represented as $\ket{v}$, while $\bra{v}$, called \textit{bra}, is its Hermitian adjoint.
Using this formalism, it comes natural, given two vectors $v$, $w$, the expressions for the inner product $\braket{v}{w}$ and the outer product $\ketbra{v}{w}$.\\
Using bra-ket notation, therefore, a qubit can be defined as
\begin{equation}
    \ket{\Psi}=\alpha\ket{0}+\beta\ket{1},
\end{equation}
where $\alpha$, $\beta$ are complex numbers satisfying the normalisation condition $|\alpha|^2+|\beta|^2=1$, and $\ket{0}=\begin{bmatrix}1\\0\end{bmatrix}$, $\ket{1}=\begin{bmatrix}0\\1\end{bmatrix}$ are two vectors, in principle arbitrary: in this case the computational basis states (connected with the logical "$0$" and "$1$") were used.\\
\begin{figure}[ht]
\centering
\includegraphics[width=.4\textwidth]{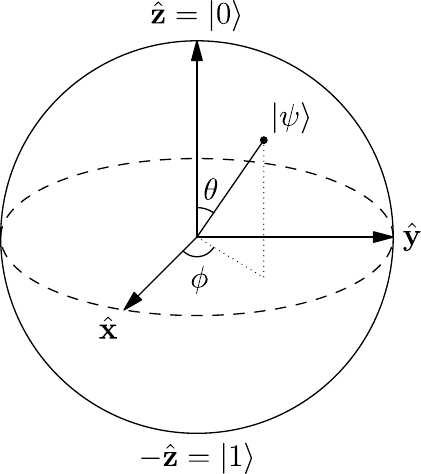}
\caption{A qubit can be described (because of the normalisation condition) by means of only two angles $\theta$, $\phi$. It can be noticed that North pole of the sphere represents the state $\ket{0}$ and South pole the state $\ket{1}$, while any other state $\ket{\Psi}$ is a superposition of the two states. Figure from~\cite{nielsen_chuang}.}
\label{fig:bloch_sphere}
\end{figure}
\par
In Figure~\ref{fig:bloch_sphere} there is a graphical representation of a qubit, commonly called Bloch sphere.
The normalisation condition guarantees that the vector has unitary magnitude and, consequently, two angles are needed to fully describe an arbitrary quantum state.\\
Therefore, a convenient way to express a quantum state is
\begin{equation}
    \ket{\Psi}=\cos\frac{\theta}{2}\ket{0}+e^{i\phi}\sin\frac{\theta}{2}\ket{1}.
\end{equation}
where $\theta,\,\phi$ are angles on the Bloch sphere and $\ket{0},\,\ket{1}$ are vectors of the computational basis.

\subsubsection*{No-cloning Theorem}
The \emph{no-cloning theorem}~\cite{wootters1982} states that it cannot exist an unitary operation able to create an identical copy of an arbitrary state.
It means that, if we have an unknown arbitrary state $\ket{\psi}_{1}\in H_1$ and a "blank" state $\ket{e}_{2}\in H_2$, with $H=H_1=H_2$, it is impossible to find an unitary transformation $U$ on $H\otimes H$ such that
\begin{equation}
    U\ket{\psi}_1\ket{e}_2=\ket{\psi}_1\ket{\psi}_2
\end{equation}
It is an important result because it negates the possibility to copy an arbitrary quantum state, differently from the classical case where information can be copied indefinitely.

\subsubsection*{Entanglement}
We already discussed that an arbitrary single qubit state can be expressed as a superposition of the basis states.
We can apply the same reasoning for a pair of qubits, but using as basis states the tensor product of the two qubits composing the system $\ket{00},\,\ket{01},\,\ket{10},\,\ket{11}$.
As a result, any arbitrary two-qubit state can be written as
\begin{equation}
    \ket{\Psi} = \alpha_{00}\ket{00} + \alpha_{01}\ket{01} + \alpha_{10}\ket{10} + \alpha_{11}\ket{11},
\end{equation}
where $\alpha_{ij}$ are normalised complex amplitudes and $\ket{ab}=\ket{ab}_{1,2}=\ket{a}_{1}\otimes\ket{b}_{2}$ is the tensor product of two different qubits $\ket{a}_{1}$ and $\ket{b}_{2}$ (for simplicity, in the tensor product indices can be omitted).
In fact, due to superposition principle, when two, or more, qubits are interacting with each other, \emph{entanglement} can occur.\\
\par
Entanglement is an notably fascinating phenomenon, and a point of strength for quantum computation.
In a formal way, a state $\ket{\Psi}$ is entangled if it cannot be expressed as a tensor product of its individual vectors $\ket{\Psi}\neq\ket{a}\otimes\ket{b}$.\\
Entanglement between qubits leads to a correlation in the physical quantities to be measured.
When one qubit is measured in a certain basis, the measurement outcomes measuring the other qubits are correlated.
Moreover, this correlation holds for any distance between the qubits.
It has been the non-locality of entanglement that was criticised by Einstein-Podolsky-Rosen in a 1935 paper (the so called EPR paradox)~\cite{EPR}, but later it was experimentally proven~\cite{Hensen2015},~\cite{Giustina2015},~\cite{Shalm2015} that quantum mechanical states are indeed non-local.\\

It is convenient to introduce some special entangled states, which are maximally entangled (in the sense of entanglement entropy).\\
\emph{Bell's states}
\begin{gather}
        \ket{\Psi^{\pm}}=\frac{\ket{00}\pm\ket{11}}{\sqrt{2}}\\
        \ket{\Phi^{\pm}}=\frac{\ket{01}\pm\ket{10}}{\sqrt{2}}
\end{gather}
\emph{Greenberger-Horne-Zeilinger (GHZ) states}
\begin{equation}
    \ket{GHZ}=\frac{\ket{0}^{\otimes N}+\ket{1}^{\otimes N}}{\sqrt{2}},\quad N>2,
\end{equation}
and \emph{W-states}
\begin{equation}
\begin{split}
    \ket{W}=\frac{1}{\sqrt{N}}&(\ket{100\dots0}+\ket{010\dots0}+\\
    &+\dotsc+\ket{000\dots1}),
\end{split}
\end{equation}
where $N$ is the number of the qubits composing the system.

\section*{Quantum Consensus}
After this not exhaustive introduction to concepts of Quantum Computation, we are approaching the main topic of our review: quantum consensus.
Because of the peculiar features of Quantum Computation (e.g. entanglement, stochastic measurement outcomes, no-cloning theorem, etc.), it is not easy or completely meaningful to define quantum consensus based on the classical counterpart.
In general, a definition of consensus for quantum systems has to take into account the fact that a quantum network, however it may be built, is intrinsically different respect to a classical network.\\
To better understand the differences and, thus, define how quantum consensus may be stated, let us discuss first about what is classical consensus.

\subsubsection*{Classical Consensus}
In classical distributed computing, when dealing with concurrent processes, state machine replication, multi-agent systems and related paradigms, the goal is to achieve an overall system reliability, despite the presence of a number of faulty processes.\\
Roughly, there are two main properties that a reliable distributed system has to satisfy:
\begin{itemize}
    \item \textbf{Liveness}: summarised as \textit{"something good will eventually occur"};
    \item \textbf{Safety}: summarised as \textit{"something bad will never happen"}.
\end{itemize}
More formally, as stated in~\cite{Mazzarella2015}, for a consensus protocol it is required that, given a common input, the state related to each subsystem $x_i(0)\in\mathbb{R}^n$ is evolving (according to certain network policies and dynamics) toward a non-trivial output state $x_i(t)$ at an arbitrary time $t\in(0,\infty)$.
The output state of each subsystem, except for a "small" set of faulty processes, converges to a configuration in which subsystems have the same output: the \emph{consensus state} $x_i(t) = x_j(t)\quad\forall i,j$.
Moreover, if the value of the consensus state is the average of the initial states, the system is said to reach an \emph{average consensus}.\\
An other way of defining consensus is to search for invariance with respect to subsystems permutations.
Given the set of all the possible permutation operations, it is sufficient to check that all the pairwise permutations $P_{\pi}$ are invariants of the system, i.e. $P_{\pi}x=x$.

\subsubsection*{Defining Quantum Consensus}
The definition of quantum consensus may be, in principle, based on its classical counterpart: even though, a classical probabilistic consensus must be considered because of the inherent stochastic nature of quantum measurement.\\
In~\cite{Mazzarella2015}, authors made a systematic study on consensus in quantum networks.
They define four classes of consensus, namely $\sigma$-expectation consensus, reduced state consensus, symmetric state consensus, and single $\sigma$-measurement consensus, are basing their definitions on symmetries and invariants of the system.
As stated by the authors, those definitions may work also with classical random variables or probability distributions of the state values, and it is worthy to point out that those definitions actually reflect the spirit of the classical consensus.
They also found hierarchies for their definitions of quantum consensus, and the implications deriving from them.
Finally, they discuss how to detect consensus in quantum networks.
They showed that studying symmetries in a quantum network is not trivial, because there is the possibility of having entanglement as correlation between qubits: due to entanglement, it becomes not obvious to check the effect of permutations on the full state.

\subsubsection*{Quantum Consensus Algorithms}
Quantum protocols and algorithms to achieve consensus over a quantum network have been proposed by researchers.
A categorisation of such algorithms leads to the identification of four categories, based on the quantum mechanical feature used to reach consensus: state invariance respect to permutations, correlations due to entangled states, state evolution by means of quantum measurements, and by means of Quantum Key Distribution (QKD) protocols.\\

\noindent\textit{Symmetric-state Consensus}\\

G.~Shi \textit{et al.} in three different papers~\cite{shi2014},~\cite{shi2015},~\cite{shi2016} present the convergence of the state of a quantum network to a consensus symmetric-state.
They use an underlying Lindblad master equation~\cite{lindblad1976} to describe the state evolution of the quantum network with continuous-time swapping operators.
Authors also prove that quantum consensus of $n$ qubits naturally defines a consensus process on an induced classical graph with $2^{2n}$ nodes.
Such a mapping allows to study quantum consensus convergence by means of already existing results for consensus on classical networks.\\

L.~Mazzarella \textit{et al.}, in~\cite{Mazzarella2015}, propose and analyse a quantum gossip-type algorithm that asymptotically converges to symmetric-state consensus states, while preserving the expectation of any permutation invariant global observable.\\

R.~Takeuchi and K.~Tsumura study distributed feedback control of quantum networks with local quantum observation and feedback~\cite{TAKEUCHI2016}.
Authors prove that a quantum consensus algorithm makes quantum states converge to a symmetric-state consensus from arbitrary initial states preserving purity; authors also show that quantum consensus algorithms can generate a W-state entanglement.\\

A work by S.~Jafarizadeh~\cite{JAFARIZADEH2016} studies the optimisation of the convergence rate of the quantum consensus algorithm over quantum network with $N$ qudits.
The model used relies on the discrete-time evolution of the consensus algorithm.
In this framework, author proves that the convergence rate of the algorithm depends on the value $d$ of the qudits.\\

F.~Ticozzi introduces two algorithms giving both an improvement to the gossip-like consensus and a new dynamic for quantum consensus~\cite{TICOZZI2016}.
The author analyses the limits of the symmetric-state consensus states, pointing out that this kind of algorithms do not reach consensus strictly speaking, but they reach consensus on the statistical properties of the variables of interest.
In this scheme, no algorithm can attain actual consensus on the output of
each local measurement.
The first proposed algorithm improves the existing gossip-type dynamics, as in general it attains a purer output state while still guaranteeing symmetric-state consensus.
The second algorithm, instead, guarantees a result closer to the idea of classical consensus: if the measurement of a local observable quantity gives a certain outcome, any subsequent measurement in all the other subsystem will return the same outcome.\\

\noindent\textit{Entanglement-based Consensus}\\

A widely discussed and studied consensus algorithm for classical networks is the one derived from a problem in distributed computing know as \emph{The Byzantine Generals Problem}~\cite{Lamport1982}.
This problem deals with the presence of faulty participants in the consensus process, with the possibility that not only those participants are defective, but they may also act deliberately against the achievement of consensus.\\
Both those problems have been studied in a quantum framework, demonstrating that quantum properties can enhance the classical results, and even solve problems otherwise believed unsolvable.\\

M.~Ben-Or and A.~Hassidim presented an algorithm for a fast quantum Byzantine agreement~\cite{ben2005}.
Authors prove that, using their algorithm, a network equipped with both quantum and classical channels can reach Byzantine agreement in $O(1)$ expected communication rounds against a strong full information, dynamic adversary,
tolerating up to the optimal $t < n/3$ faulty players in the synchronous setting, and up to $t < n/4$ faulty players for asynchronous systems.\\

L.~Helm proposed an algorithm solving the same problem~\cite{helm2008}, with a particular focus on addressing the FLP impossibility of distributed consensus in an asynchronous setting~\cite{fischer1985}.
In particular, author claims that his algorithm can solve the consensus problem in a completely asynchronous setting, without the need of classical communication.
In~\cite{golab2020}, authors question if is it possible to reach consensus using the protocol by~\cite{helm2008}, since that quantum algorithm cannot provide deterministic agreement and validity.\\



\noindent\textit{Measurement-based Consensus}\\

As prescribed by the postulates of quantum mechanics, the operation of measuring a quantum state determines the collapse of the quantum state in an eigenstate of the operator involved into the measurement.
Loosely speaking, this means that a quantum state, when it has been measured, evolves into a new state determined by the measurement itself.\\

With this premise, consensus for a quantum hybrid network model is proposed and analysed in~\cite{shi2017}. G.~Shi \textit{et al.} are considering, as a quantum hybrid network, a network consisting of a number of nodes each holding a qubit, while communication for reaching consensus is performed over classical channels.
The proposed protocol drives such a quantum hybrid network to a consensus in the sense that all qubits reach a common state.
This is achieved performing measurements on qubits, thus outcomes are exchanged between nodes by means of messages via classical communication channels.
In order to address the issues connected to the large amount of messages in a centralised solution, authors also develop a distributed Pairwise Qubit Projection (PQP) algorithm, demonstrating that a quantum hybrid network almost surely converges to a consensus status for each qubit.\\

\noindent\textit{QKD-based Consensus}\\

X.~Sun \textit{et al.} proposed a quantum communication protocol to achieve Byzantine agreement among multiple parties without entanglement~\cite{sun2020}.
The protocol relies on the unconditional security given by Quantum Key Distribution (QKD).
Sequences of correlated numbers shared between semi-honest distributors of quantum keys is the solution on which authors base their proposal.
Further development of this protocol could consider a low-dimensional entanglement to replace the need of semi-honest participants to the key distribution in the protocol.


\section*{Discussion}
In this paper we present an overview on quantum consensus, i.e. reaching agreement in quantum networks.
Our goal is to introduce readers with no knowledge about quantum mechanics to the state of the art on quantum consensus.
We gave some crude tools to understand quantum formalism important in quantum information and computing.
Then, we reported the literature on the topic, dividing the discussion in two main subtopics: definition of quantum consensus and proposed protocols.
In presenting the proposed protocols, we classified them into four groups depending on the feature used to reach consensus, namely: symmetric-state consensus, entanglement-based consensus, measurement-based Consensus, and QKD-based consensus.\\

It is important to notice that the literature about quantum consensus is still scarce, especially if compared to the amount of works on quantum distributed systems and networks.\\

Further study may be related to the application of consensus in the (proposed) quantum distributed ledgers, prototypes of the quantum blockchains.

\bibliography{biblio}{}
\bibliographystyle{unsrt}

\end{document}